\documentclass[preprint,compress,3p]{elsarticle}
\usepackage{amsmath}
\usepackage{amsfonts}
\usepackage{amssymb}
\usepackage{float}
\usepackage{subfig}
\usepackage{textcomp}

\usepackage{cleveref}

\usepackage[usenames,dvipsnames]{color,colortbl}

\usepackage{graphicx}
\usepackage[export]{adjustbox}[2011/08/13]

\begin{document}
\begin{frontmatter}
\journal{International Journal of Engineering Science}
\title{Poroelasticity in the presence of active fluids}
\author[1,2]{R. Cavuoto}
\author[1,3]{S. Scala}
\author[2]{A. Cutolo}
\author[3]{G. Mensitieri}
\author[2]{M. Fraldi\corref{cor}}
\ead{fraldi@unina.it}

\address[1]{Department of Neurosciences, Reproductive sciences and Dentistry, University of Naples Federico II, Naples, Italy}
\address[2]{Department of Structures for Engineering and Architecture, University of Naples Federico II, Naples, Italy}
\address[3]{Department of Chemical, Material and Production Engineering, University of Naples, Federico II, Naples, Italy}
\cortext[cor]{Corresponding authors}

\begin{abstract}
    This work presents a model for characterizing porous, deformable media embedded with magnetorheological fluids (MRFs). These active fluids exhibit tunable mechanical and rheological properties that can be controlled through the application of a magnetic field, which induces a phase transition from a liquid to a solid-like state. This transition profoundly affects both stress transmission and fluid flow within the composite, leading to a behaviour governed by a well-defined threshold that depends on the ratio between the pore size and the characteristic size of clusters of magnetic particles, and can be triggered by adjusting the magnetic field intensity. These effects were confirmed through an experimental campaign conducted on a prototype composite obtained by imbibing a selected MRF into commercial sponges.\\
    To design and optimize this new class of materials, a linear poroelastic formulation is proposed and validated through comparison with experimental results. The constitutive relationships, i.e. overall elastic constitutive tensor and permeability, of the model are updated from phenomenological observations, exploiting the experimental data obtained for both the pure fluid and the composite material. The findings demonstrate that the proposed simplified formulation is sufficiently robust to predict and optimize the behaviour of porous media containing MRFs. Such materials hold significant promise for a wide range of engineering applications, including adaptive exosuits for human tissue and joint rehabilitation, as well as innovative structural systems.
\end{abstract}

\end{frontmatter}

\section{Introduction}
\label{Introduction}
The rehabilitation of human tissues and joints, particularly following injuries or degenerative conditions, demands innovative solutions that are both lightweight and minimally invasive. Traditional exoskeletons \cite{facuciello2024}, while effective in providing support and enhancing mobility, often suffer from drawbacks such as bulkiness, rigidity, and lack of adaptability to individual patient needs \cite{siviy2022,Yang2024,marconi2019}. These limitations can impede the recovery process and limit the effectiveness of rehabilitation protocols. Consequently, there is a growing interest among the scientific community to develop exosuits that are not only lighter and more comfortable but also capable of offering customizable support tailored to the unique biomechanical requirements of each patient.\\
Magnetorheological (MR) fluids, consisting of micron- and nano-sized magnetic particles suspended in a carrier liquid with additives, have emerged as a promising component in adaptive materials. 
The main characteristic MRF is that when subjected to an external magnetic field, these fluids exhibit a transition from a liquid-like to a solid-like state, characterized by a dramatic increase in yield stress. Their mechanical response is therefore variable but still retains some key features, the most striking one being the presence of a threshold of the applied stress that is required to initiate flow. This behaviour is typical of Bingham plastic fluids \cite{bingham1916,bingham1922,bingham1992,obrien2014,moller2017,martinez2020,moatimid2024,gao2024}, which flow like a viscous fluid when subjected to stresses exceeding a critical yield value but behave like a solid below this threshold. MR fluids, thus, share this dual nature, while furthermore having the ability to display a tunable transition of properties in real-time via the applied magnetic field. Their ability to transition between flowing and solid-like states, makes MR fluids particularly suitable for applications where controllable mechanical properties are crucial.\\
A promising approach to achieving this goal lies in the integration of porous deformable media with plastic fluids \cite{nash2017,carrillo2019,gunawan2022,talon2024}, and in particular with magnetorheological (MR) fluids \cite{jackson2018,bodniewicz2018}. The resulting material is typically called \textit{Magnetorheological composite}. Porous media, known for their lightweight properties, provide a structural matrix that can (i) limit the phenomenon of dissipated particle sedimentation of MRF, (ii) maintain the fluid in a controlled space, and in turn (iii) be dynamically adjusted when combined with MR fluids, which consist of micron- or nano-sized magnetic particles suspended in a carrier fluid, exhibiting a remarkable ability to change their rheological properties in response to an applied magnetic field. This tunability allows for precise control over the stiffness and damping characteristics of the material, making it particularly suitable for applications requiring variable mechanical properties, and for instance, providing the optimal level of support and resistance throughout different stages of rehabilitation.\\
To fully harness the potential of this approach, we put forward a poroelastic model for porous deformable media embedded with MR fluid. As an initial step, the linear response of the poroelastic solid is analyzed. Within the framework of linear poroelasticity, we leverage the well-established theory of three-dimensional consolidation by Biot \cite{BIOT1941,coussy2004,Wang2000}, extending its applicability to the current case study by treating the interstitial fluid as a viscous medium whose mechanical response is influenced by an external agent (magnetic field). When this field is deactivated, the behaviour mirrors that of a classical poroelastic model. However, when the magnetic field is activated, the fluid's characteristics swiftly shift towards a more solid-like response. Consequently, the constitutive relationships of poroelasticity are modified so that the fluid's contribution to equilibrium gradually diminishes, while the elastic response of the solid is enhanced by the presence of the new phase. This theoretical and practical framework has been scarcely investigated in the literature \cite{thekkethil2024}.\\
Section \ref{ch:magnetic_fluid} provides a brief introduction to MRF fluids, emphasizing the key aspects of their mechanical response, which is subsequently integrated into the poroelastic model. As established in the literature, the behaviour of these smart fluids can vary depending on their composition and external conditions, with factors such as the type of carrier fluid, choice of additives, and the size and shape of magnetic particles playing significant roles. Special attention is given to a specific MRF fluid synthesized in the laboratory of the University of Naples Federico II, which offers a good qualitative representation of their behaviour.\\
Section \ref{ch:sponges} presents the mechanical characterization of a commercial sponge that has been used to build, in the following Section \ref{ch:composite}, the composite sponge filled with MRF then tested using an ad hoc 3D printed apparatus for oedometric tests.\\
In Section \ref{ch:model}, the linear formulation of poroelasticity is succinctly presented to familiarize the reader with the case simplification and the chosen notation, ultimately deriving the constitutive relationship for linear poroelasticity. These relationships are further elaborated in subsections \ref{ch:constitutive_porous} and \ref{ch:fluid_constitutive}, where the elastic porous solid and the fluid are detailed to account for the system's evolution under the influence of the magnetic field. Thermodynamic consistency is addressed and ensured through the definition of a \textit{complementary condition}, which mechanically expresses the fluid's transition. Furthermore, the balances of linear momentum and of fluid content are presented in the same Section, subsections \ref{ch:balance_solid} and \ref{ch:balance_fluid}, respectively.\\
Finally, the model is tested against the experimental results obtained for the specific problem of the oedometric test — a three-dimensional setup inducing one-dimensional loading and evolution conditions, Section \ref{ch:benchmark_1}, and in Section \ref{ch:benchmark_2} the ability of the model to reproduce spatial tuneability of the magnetic field and thus of the mechanical properties is verified by means of a numerical experiment of a beam subjected to benging loading conditions.\\
Through this model, the Authors aim to capture the intricate interactions between the porous matrix and the MR fluid, considering factors such as fluid flow, magnetic field application, and mechanical deformation.

In the following, we use regular uppercases and lowercases ($U$,$S$,$\phi$,$\mu$,...) for scalar fields, bold lowercases ($\mathbf{q}$,$\mathbf{m}$,...) for vectors, bold uppercases ($\mathbf{T}$,$\mathbf{E}$,$\mathbf{K}$,...) for second-order tensors and blackboard-bold ($\mathbb{C}$) for fourth-order tensors. As has been explained above, the present study puts forward a model for poroelastic solids with special fluids embedded in them, therefore, in the following, material properties are defined for the different constituents and the overall solid: the properties of the material comprising the matrix in which the fluid flows are indicated through the index \lq \lq mat"; the properties of the fluid are indicated through \lq \lq F"; the properties of the overall solid in drained conditions with no subscript; lastly, the properties of the overall solid in undrained conditions with the apex $(u)$.

\section{Magnetorheological fluids}
\label{ch:magnetic_fluid}
\begin{figure}
    \centering
    \includegraphics[width=\linewidth]{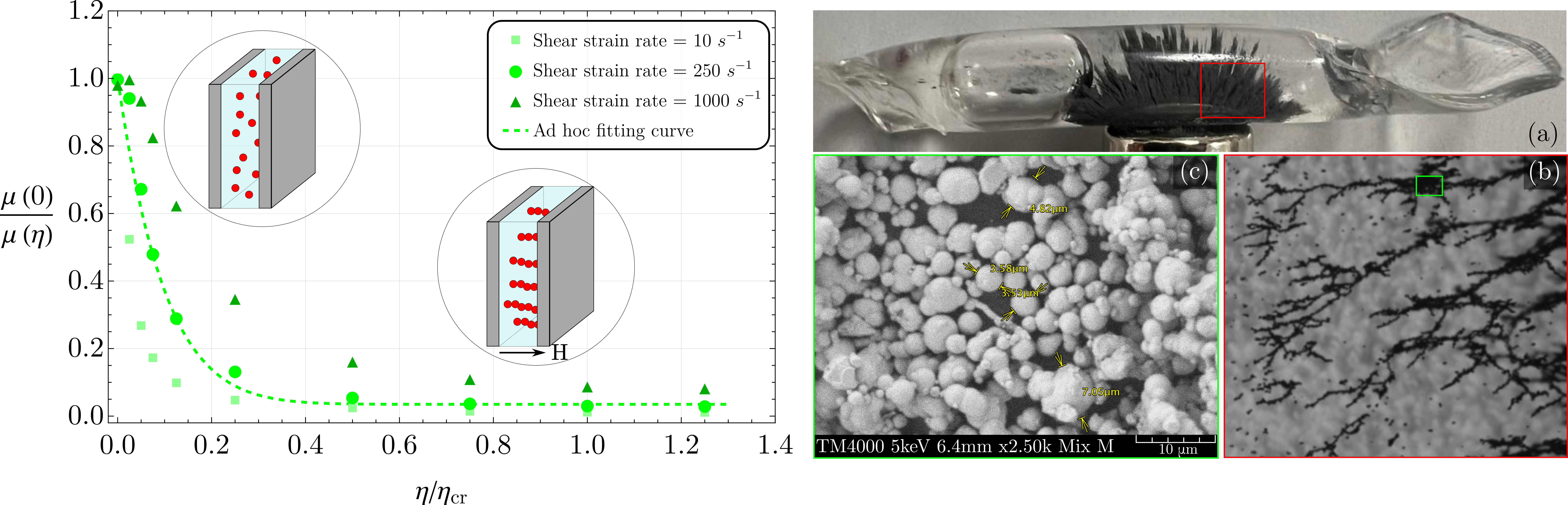}
    \caption{On the left: viscosity measurements of a specific magnetorheological fluid (MRF). The fluid is characterized by: a base silicon oil with a viscosity of 100 cSt; $3\%$wt hydrophilic Fumed FS(HL) as hydrophilic additive; $20\%$wt of carbonyl iron (CI) powder in the form of microparticles with an average diameter of 8~{\textmu}m.  Measurements were performed with a controlled-stress Physica MCR 702 Rheometer (Anton Paar, Graz, Austria), equipped with the magnetorheological device of Physica (MRD 70/1T). A plate-plate geometry with plates of 20 mm diameter and a gap of 0,5 mm was used. A homogeneous magnetic field with different strengths (0, 10, 25, 40, 60, 80, 170, 350, 510, 670, 800 mT), perpendicular to the shear flow direction, was applied, and have then been normalized with respect to a reference value $\mu(0)$ which depends on the shear rate of the test: $\mu(0)=0.86\textup{Pa}\ \textup{s}$, $\mu(0)=1.01\textup{Pa}\ \textup{s}$ and $\mu(0)=3.36\textup{Pa}\ \textup{s}$ for a shear rate of $\dot{\gamma}=1000\ s^{-1}$, $\dot{\gamma}=250\ s^{-1}$ and $\dot{\gamma}=10\ s^{-1}$, respectively. The magnetic field has been normalized with respect to a value, called $\eta_{cr}$ in the following that, as will be explained in the text, emerges from the interaction of the fluid with the microstructure of the solid in which it is embedded. The ad hoc fitting curve in the graph is obtained from the expression of the viscosity in equation (\ref{eq:permeability}). On the right: (a) the analyzed MRF upon activation of an external magnetic field builds internal structures made by the metallic particles that (b) assemble along the direction of the magnetic field with shape and length that depend on (c) the micro particle's size and magnetization properties.}
    \label{fig:mrf_viscous}
\end{figure}

Magnetorheological fluids \cite{rabinow1948,china2019,kwon2017,Kumar2019,Thiagarajan2021} are smart fluids typically made of three components: a carrier liquid (mainly mineral oils), ferromagnetic particles and additives. The scope of the presence of these particles is to actively manipulate and control the viscosity of the mixture. In fact, when subjected to a magnetic field, the magnetic particles, in the form of micron or nanoparticles, magnetize and align along the direction of the magnetic field, creating structures that alter the carrier fluid's viscous response by one or two orders of magnitudes, see Fig. \ref{fig:mrf_viscous}. Their versatility \cite{jolly1998} is due to the rapid field responsiveness, noiseless operation, relative insensitivity to small quantities of dust or contaminants, and ease of control.\\
The choice of the synthetic oil \cite{jiang2011}, of the additive \cite{aruna2021}, and size distribution of the ferromagnetic particles \cite{Li&Tian2019}, are all parameters that greatly impact the material mechanical response. This impact is measured in terms of sedimentation ratio, prevention of the oxidation of the CI particles, the maximum increase in viscosity registered under magnetic induction, and also in terms of magnetic saturation, coercivity and retentivity, all main features of the magnetic hysteretic loop. Apart from the fluid intrinsic characteristics, the shear strain rate acting on the material also plays a relevant role.\\
The magnetization of the MRF affects more than just the rheological properties of the overall mixture. It is, in fact, evident how the behaviour of the fluid resembles more and more that of a solid as the magnetization reaches saturation, implying that the static behaviour is just as much modified. The reason behind this observation lies in the assemblage of the microparticles. In fact, close to magnetic saturation, the particles form clusters very rapidly, and the chains assemble in large blocks.

The present study aims at developing an augmented theory of poroelasticity that can be easily tuned to satisfy the specific material response of the magnetorheological composite. Therefore, as a preliminary analysis, only a qualitative viscosity curve (see Fig. \ref{fig:mrf_viscous}) is chosen. 

\section{Porous elastic deformable matrix} \label{ch:sponges}
\begin{figure}
    \centering
    \includegraphics[width=\linewidth]{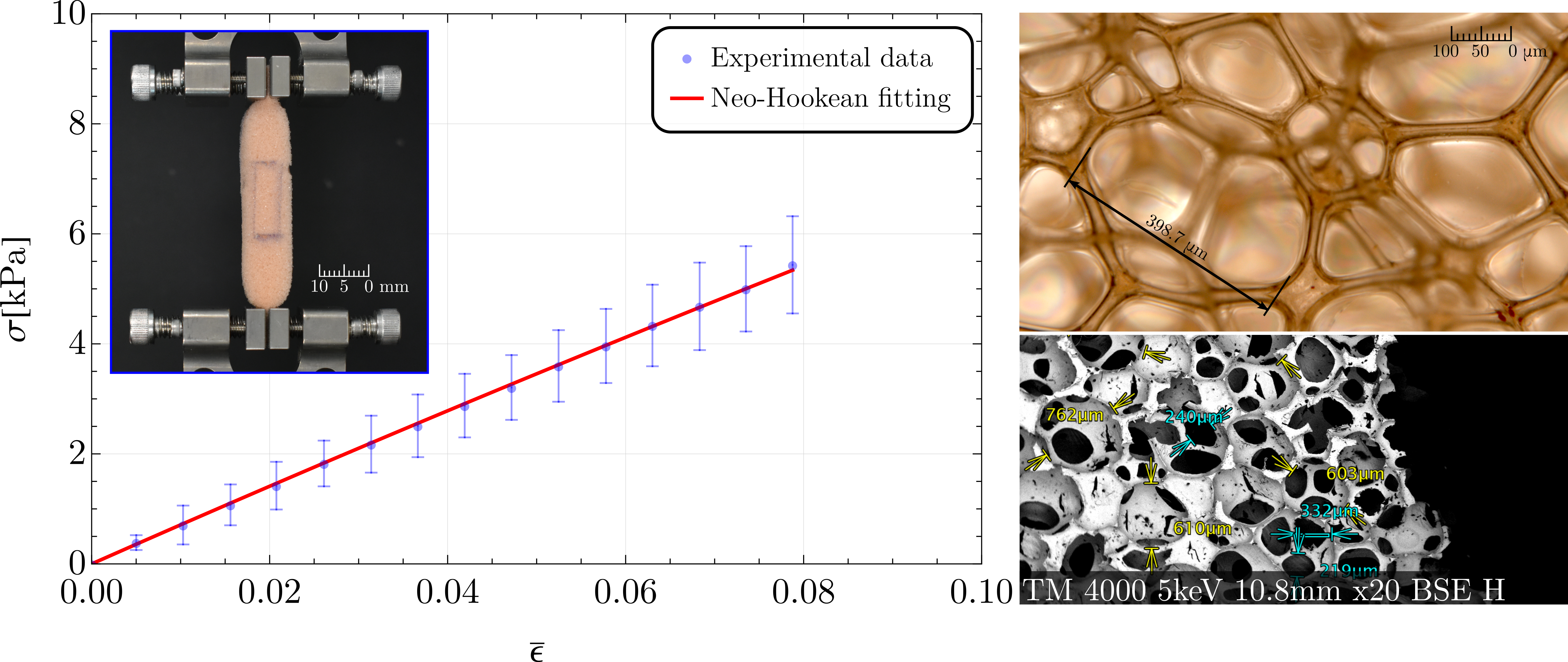}
    \caption{Tensile test carried out on the chosen sponge samples. On the left, average stress vs average strain measured during the test: the dots report the actual experimental measurements, while the continuous red line is obtained through a Neo-Hookean fitting (see Appendix for more details). On the right, an optical microscopy (above) and a scanning electron microscopy (below) displaying the partially open porosity with characteristic void size.}
    \label{fig:mech_sponge}
\end{figure}

The analytical formulation presented in the following is tested against experimental measurements on prototypes of solid sponges with MRF embedded in them. An experimental campaign has been conducted on commercial porous sponges that have been characterized through standard mechanical testing. Thus, an experimental campaign has been conducted on a selected set of commercial sponges whose mechanical response to an uniaxial tensile stress is depicted in Figure \ref{fig:mech_sponge}. 
From the tests, the sponges have shown a linear elastic isotropic response. To completely define the two elastic constants of such sponges, an oedometric test has been carried out. From the knowledge of the oedometric modulus ($E_{oe}$) and Young's Modulus ($E$), Poisson ratio ($\nu$) of the material has been derived: $E=73.8 \pm 7.00\ \textup{kPa}$, and $\nu = 0.248 \pm 0.003 $ .\\
Furthermore, to investigate the internal structure of these sponges, visual investigations have been performed aided by optical microscopy and scanning electron microscopy. Optical microscopy has shown a translucent, partially open, and intricate structure of pores resulting from a trabecular-like skeleton structure with pores of exagonal shape and a maximum diameter of around 400 $\mu$m. The scanning electron microscope has highlighted that this porosity tends to be hierarchical, with macropores hosting a varying number of smaller pores (from 1 to 3) of approximately half the diameter of larger ones. The hexagonal symmetry of these pores with three smaller voids (see Figure \ref{fig:mech_sponge}, bottom-right picture) is clearly one of the factors responsible for the overall macroscopic isotropy that emerged from the mechanical testing, the other one being the random distribution of these cavities in the specimen.\\
Overall porosity has been evaluated by means of dry and wet weight comparison. The wet evaluation has been performed by imbibition of the sponge by a 10cp silicone oil. Through the difference in weight between the two conditions and by knowing the specific weight of the oil the porosity was estimated to be around $83\%$.

\section{MRF composite sponge} \label{ch:composite}
Building upon the results presented in the previous sections—namely, the formulation of the magnetorheological fluid and the selection of a suitable porous matrix—we proceeded to integrate the two components in order to obtain the final composite material. To this end, cubic specimens of the porous sponge were carefully cut, each with an edge length of 2.49 cm, and subsequently impregnated with the magnetorheological fluid until the pore volume was fully saturated.\\
The imbibition process was carried out under gravity, allowing the fluid to flow from the top after being poured onto the upper surface of the cubic sponge specimen. The low surface tension of the carrier oil facilitated the rapid infiltration and percolation of the MRF throughout the porous network. In addition, mechanical stimuli in the form of small compressive cyclic loads were applied to promote fluid movement and to ensure complete penetration of the oil into all cavities.\\
The effectiveness of the impregnation process was verified through both optical microscopy and CT scanning (ZEISS METROTOM 6 scout - GOM CT), which confirmed that the fluid had homogeneously filled the internal porosity of the matrix. These analyses provided confidence that the composite retained the structural characteristics of the sponge while incorporating the functional properties of the magnetorheological fluid.\\
Finally, to investigate the mechanical response of the material under controlled conditions, oedometric compression tests were conducted. The test consists of a transversely confined compression applied using a porous plate to allow fluids to escape the inner volume.
For this purpose, a dedicated test apparatus was designed and manufactured using 3D printing, ensuring that both the geometric and physical requirements of the composite specimens were adequately met.\\
The testing apparatus for the oedometric experiment consisted of a cubic box made of 3D-printed resin. The internal housing of the box was shaped as a cube with an edge length of $L=2.49 \pm 0.01$ cm, designed to accommodate the saturated specimen. The box walls were 0.5 cm thick; however, dedicated recesses were incorporated on the free sides of the cube to allow the insertion of circular magnets. In this configuration, the magnets could be positioned at a distance from the specimen smaller than the nominal 0.5 cm wall thickness of the oedometer, thereby maximizing the influence of the applied magnetic field on the sample.\\
The upper side of the oedometer was closed by inserting a porous plate with holes of 1 mm radius, allowing the MRF fluid to drain and thereby minimizing the pressure fluctuations associated with the filtration. The specimens were then tested by displacing (therefore in a displacement-induced fashion) the oedometer head using a DMA (TA instruments, ElectroForce\textsuperscript{\tiny{\circledR}} 3300  Series III), with a velocity of the moving head of 0.025mm$/$s up to a value of approximately 7 mm corresponding to an average value of deformation of $\epsilon_{max}\approx0.28$.
\begin{figure}[H]
    \centering
    \includegraphics[width=\linewidth]{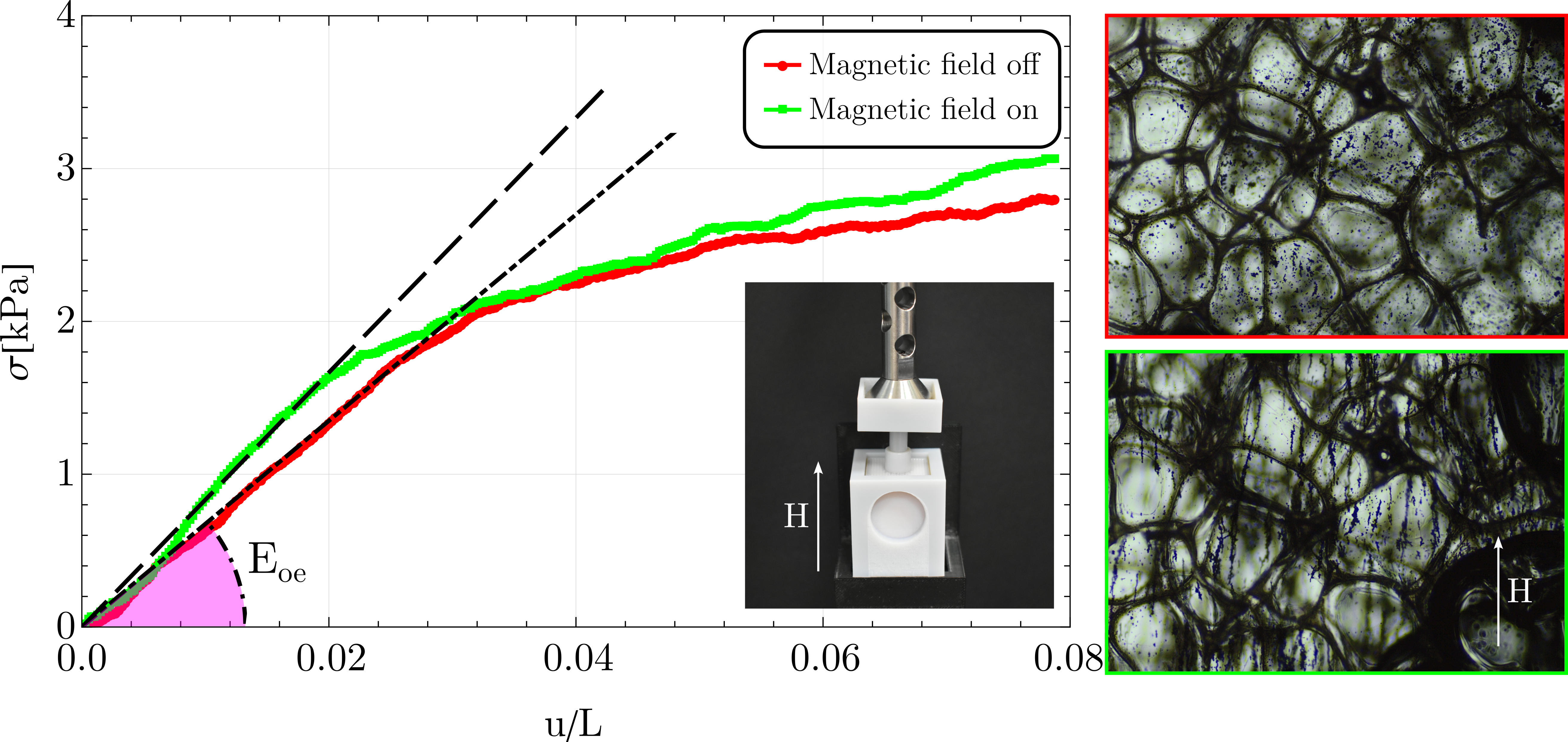}
    \caption{Mechanical response of the deformable sponge embedded with MRF (10cP, 20$\%$ wt of CI) tested in oedometric conditions using the ad hoc prepared set-up. The magnetic field is turned off at first (red lines), and then turned on (green lines). The stiffness exhibited by the specimen in such conditions corresponds to the so-called oedoetric modulus $E_{oe}=\lambda+2G$ ($\lambda$ and $G$ being the two Lamè constants) of the deformable porous sponge, which are typically referred to in poroelasticity and the dry elastic properties. The magnetic field $H$, when present, is applied by means of a thin cylindrical neodymium magnet placed beneath the oedometer in between the lower walls and the anchoring to the testing machine. As such, the magnetic field is not everywhere parallel to the loading direction but exhibits classical curved paths in space. On the right, optical microscopy shows the variation in particles' direction and alignment when a magnetic field is applied upon the composite sponge-MRF.}
    \label{fig:placeholder}
\end{figure}
The tests were carried out both in the absence and in the presence of a magnetic field. The magnetic field was applied by inserting a circular neodymium magnet with a diameter of approximately 2 cm into the dedicated recess of the oedometer. When the field is activated, the ferromagnetic particles dispersed in the MRF reorient and align into rigid chains along the direction of the magnetic field. For these chains to effectively contribute to the composite response, they must be aligned along the loading direction of the solid. Conversely, if the magnet is positioned transversely (i.e., within the lateral compartments of the oedometer), the particle chains are oriented along a direction of zero strain, and the contribution of the fluid to the overall mechanical response becomes negligible \cite{jackson2018}.

\section{Porous solids hosting a viscous fluid: poroelastic formulation}
\label{ch:model}
The theory of poroelasticity \cite{BIOT1941,coussy2004,Wang2000} is a macroscopic (in the sense of homogenized) formulation of mechanical coupling between porous elastic solids and the fluids permeating them.\\
Following \cite{coussy2004}, for an isothermal process, the Clausius-Plank inequality specialises in
\begin{equation*}
    \mathbf{T}:\dot{\mathbf{E}}+(p-p_0)\dot{\phi}-\dot{\psi}-\nabla \left( p-p_0\right)\cdot \mathbf{q}\geq 0\quad ,
\end{equation*}
where $\mathbf{T}:\mathbf{E}$ is an energy-conjugated pair composed by the second Piola-Kirchhoff stress tensor and Cauchy-Green strain tensor, $p$ and $p_0$ are the pressure of the pores and its initial -- spatially uniform -- value, $\phi$ is the fluid volume fraction, $\psi$ is Helmholtz free energy density of the porous solid, $\rho_f$ is the true density of the fluid -- fluid mass divided by fluid volume -- and $\mathbf{q}$ is the fluid flux.\\
Helmholtz free energy  is postulated to be a function of the deformation of the porous solid (homogenized solid), $\mathbf{E}$, and pressure in the pores, $p$, making its material time derivative as
\begin{equation*}
    \mathbf{T}:\dot{\mathbf{E}}+(p-p_0)\dot{\phi}-\nabla (p-p_0)\cdot \mathbf{q} \geq  \frac{\partial \psi}{\partial \mathbf{E}}:\dot{\mathbf{E}}+\frac{\partial \psi}{\partial \phi}\dot{\phi}_F \quad .
\end{equation*}
Further development leads to:
\begin{equation}
    \left(\mathbf{T}-\frac{\partial \psi}{\partial \mathbf{E}} \right):\dot{\mathbf{E}}+\left(p-p_0-\frac{\partial \psi}{\partial \phi}\right) \dot{\phi}_F-\nabla \left( p-p_0\right)\cdot \mathbf{q}\geq 0\quad .\nonumber
\end{equation}
Given the arbitrariness of the increments, a possible position is clearly
\begin{equation}
    \mathbf{T} = \frac{\partial \psi}{\partial \mathbf{E}}\quad , \quad p=p_0+\frac{\partial \psi}{\partial \phi}\quad .
    \label{eq:constitutive_T_p}
\end{equation}
Lastly, a Darcian-type of filtration in the porous solid,
$\mathbf{q}=-\mu(\eta)^{-1}\ \mathbf{K}\nabla p$,
implies
$\nabla p\cdot \mathbf{K}\nabla p\geq 0$.
The latter condition is satisfied for a positive definite second-order tensor $\mathbf{K}$, which indeed is the case for permeability. As previously mentioned, $\eta$ represents a static effect of the viscosity of the fluid onto the solid skeleton and is assumed to depend on the magnetization of the magnetorheological fluid $\mathbf{m}$.
Now, perform the Legendre transform to change variables
\begin{equation*}
\Psi(\mathbf{E},p)=\psi(\mathbf{E},\phi)-\phi\ (p-p_0)\quad.
\end{equation*}
Using this transformation, the incremental energy is reformulated as
\begin{equation*}
    \dot{\Psi}=\dot\psi-\dot{\phi}_F(p-p_0)-\phi\dot{p}
\end{equation*}
or by performing the time differentiation
\begin{equation*}
    \frac{\partial \Psi}{\partial \mathbf{E}}:\dot{\mathbf{E}}+\frac{\partial \Psi}{\partial p} \dot{p}=\frac{\partial \psi}{\partial \mathbf{E}}:\dot{\mathbf{E}}+\frac{\partial \psi}{\partial \phi}\dot{\phi}_F-\dot{\phi}_F(p-p_0)-\phi\dot{p} \nonumber
\end{equation*}
or
\begin{equation*}
    \left(\frac{\partial \Psi}{\partial \mathbf{E}}-\frac{\partial \psi}{\partial \mathbf{E}}\right):\dot{\mathbf{E}}+\left(\frac{\partial \Psi}{\partial p}+\phi\right)\dot{p}=\left(\frac{\partial \psi}{\partial \phi}-(p-p_0)\right)\dot{\phi}_F
\end{equation*}
which by means of eq. (\ref{eq:constitutive_T_p}) and by regrouping terms one has
\begin{equation}
    \frac{\partial \Psi}{\partial \mathbf{E}}=\frac{\partial \psi}{\partial \mathbf{E}}\quad , \quad \frac{\partial \Psi}{\partial p}=-\phi\quad .
    \label{eq:constitutive_T_phi}
\end{equation}
Equation (\ref{eq:constitutive_T_phi}) is implicitly stating that the new variable is a function of the unknown fields $(\mathbf{E},p)$. Thus, linearization of the fluid fraction allows to write:
\begin{equation*}
    \phi(\mathbf{E},p)=\phi_{F,0}+\frac{\partial \phi}{\partial \mathbf{E}}:\mathbf{E}+\frac{\partial \phi}{\partial p}(p-p_0)=\phi_{F,0}-\frac{\partial^2 \Psi}{\partial \mathbf{E}\partial p}:\mathbf{E}-\frac{\partial^2 \Psi}{\partial p^2}(p-p_0)
    \label{eq:constitutive_pores}
\end{equation*}
or through the introduction of the Biot coefficient $M=-\partial ^2\Psi/\partial p^2$ and effective stress tensor $\mathbf{A}=-\partial ^2\Psi/\partial \mathbf{E}\partial p$, one has:
\begin{equation}
    \phi(\mathbf{E},p)=\phi_{F,0}+\mathbf{A}:\mathbf{E}+M^{-1}(p-p_0)\quad .
    \label{eq:poroelastic_pores}
\end{equation}
If one now plugs the constitutive relationship just derived into (\ref{eq:constitutive_T_phi}) a simple integration with respect to the pressure ($p$) immediately shows
\begin{equation*}
    \Psi=\overline{\Psi}(\mathbf{E})-(p-p_0)\phi_{F,0}-(p-p_0)\mathbf{A}:\mathbf{E}-\frac{1}{2M}(p-p_0)^2\quad ,
\end{equation*}
where the energy term associated with the elastic deformations of the solid is, in a linear fashion, the quadratic St.Venant-Kirchhoff strain energy density. This energy is related to the overall elastic properties of the porous solid and thus, even if the isotropy of the skeleton can be assumed with no great effect on the characterization, is affected by the distribution of the pores. If isotropy is still retained at the macroscopic level, due to a random and homogeneous distribution of the pores in the matrix, then
\begin{equation*}
    \overline{\Psi}=\frac{1}{2}\mathbf{E}:\mathbb{C}:\mathbf{E}\quad ,
\end{equation*}
being $\mathbb{C}$ the fourth-order elasticity tensor of the homogenized material. To account for the effects of viscosity of the fluid onto the matrix through the shear stresses exchanged between the two continua, a modification of the constitutive tensor is necessary, and is treated in Section \ref{ch:constitutive_porous}.\\
Hence, relation (\ref{eq:constitutive_T_phi}) on the left, one has
\begin{equation}
    \mathbf{T}=\mathbb{C}:\mathbf{E}-(p-p_0)\mathbf{A}\quad ,
    \label{eq:poroelastic_stress}
\end{equation}
which is Terzaghi statement on effective stress $\mathbf{T}^{\textup{{eff}}}=\mathbb{C}:\mathbf{E}=\mathbf{T}+\mathbf{A}(p-p_0)$ \cite{BIOT1941,COWIN2007}. Hence, the constitutive analysis of linear poroelasticity allows us to write the strain tensor as
\begin{equation*}
    \mathbf{E}=\mathbb{S}:\mathbf{T}^{\textup{eff}}\quad ,
\end{equation*}
where $\mathbb{S}=\mathbb{C}^{-1}$ is the drained compliance fourth-order elasticity tensor. The Biot effective stress coefficient tensor, $\mathbf{A}$, is symmetric and in linear poroelasticity is commonly assumed to be equal to $\mathbf{A}=(\mathbb{I}-\mathbb{C}\ \mathbb{S}_{\textup{mat}}):\mathbf{I}$, in which $\mathbb{S}^{\textup{mat}}$ is the compliance fourth-order tensor of the matrix.\\
In the following, the constitutive relationships of poroelasticity just derived \cref{eq:poroelastic_stress,eq:poroelastic_pores}, are particularized for the case of study.

\subsection{Constitutive law of the porous solid} \label{ch:constitutive_porous}
In a homogenized fashion proper of poroelasticity, the effective elastic properties represented by the elasticity tensor $\mathbb{C}$ \cite{COWIN2007}, depend on the elastic properties of the matrix (two for isotropy) and on parameters representing the microstructure organization, typically the total porosity $\phi$, pores size and shape $g$, and lastly due to the participation of the fluid also on the fluid viscosity or, in the case of MRF fluids, on magnetization ($\eta$) since in this case the two quantities are directly related. Accordingly, thus $\mathbb{C}=\mathbb{C}(E_{\textup{mat}},\nu_{\textup{mat}},\phi,g; \eta)$. By invoking a customary approach, that of continuum Mixture Theory, one can write:
\begin{equation}
    \mathbb{C}=(1-\phi)^{\beta}\mathbb{C}_{\textup{mat}}+\phi^{\beta}f(\eta) \mathbb{C}_{\textup{F}}^O\quad.
    \label{eq:overall_elastic_tensor}
\end{equation}
where in an isotropic linear elastic setting $\mathbb{C}_{\textup{mat}}$, the constitutive tensor of the solid skeleton, and $\mathbb{C}_{\textup{F}}^O$, the constitutive tensor of the elastic fluid have a structure of the kind: $\lambda_i\mathbf{I}\otimes \mathbf{I}+2G_i\mathbb{I}$, implying:
\begin{equation}
    K=(1-\phi)^{\beta}K_{\textup{mat}}+\phi^{\beta}f(\eta)K_F^O\quad , \quad G=(1-\phi)^{\beta}G_{\textup{mat}}+\phi^{\beta}f(\eta)G_F^O \quad ,
    \label{eq:constitutive_scaling}
\end{equation}
where $K$ and $G$ are the overall bulk and shear moduli, while $K_F^O$ and $G_F^O$ are the bulk and shear moduli of the fluid when the magnetic field is off. 
To retain positive definiteness of the constitutive tensor the function $f(\eta)$, representing phase transition of the fluid, is restricted to being a positive scalar function. To reproduce the abrupt and fast phase transition of MRF one can choose:
\begin{equation}
    f(\eta)=H(\eta-\eta_{\textup{cr}}) \approx \frac{1}{1+e^{2s(1-\eta/\eta_{cr})}}\quad ,
    \label{eq:f}
\end{equation}
where $H$ is the Heaviside function, $\eta_{cr}$ is a critical threshold whose physical meaning is explained in the following, and $s$ is a parameter that regulates the slope of the function $f$.\\
Definition (\ref{eq:overall_elastic_tensor}), an assumption that simplifies the complex microstructural interaction between the various mechanical and geometrical parameters listed above in a relatively common way, reproduces the physical characteristics of the system at hand, so that whenever magnetization $\eta$ is null, the fluid can flow with a viscosity that is unchanged. Upon activation of the magnetic field, namely for a positive value of magnetization $\eta$, the fluid viscosity increases but flow can still occur as in a normal porous elastic matrix. As explained, the increase in viscosity of a MRF is a consequence of the formation of chains of magnetic particles that form continuously at low level of magnetization working as a skeleton for the fluid. Further increase of the magnetic field leads to the formation of larger and larger cluster of particles until at a certain threshold, flow is impeded by these clusters obstructing the channels in the porous matrix. 
\begin{figure}[H]
    \centering
    \includegraphics[width=\linewidth]{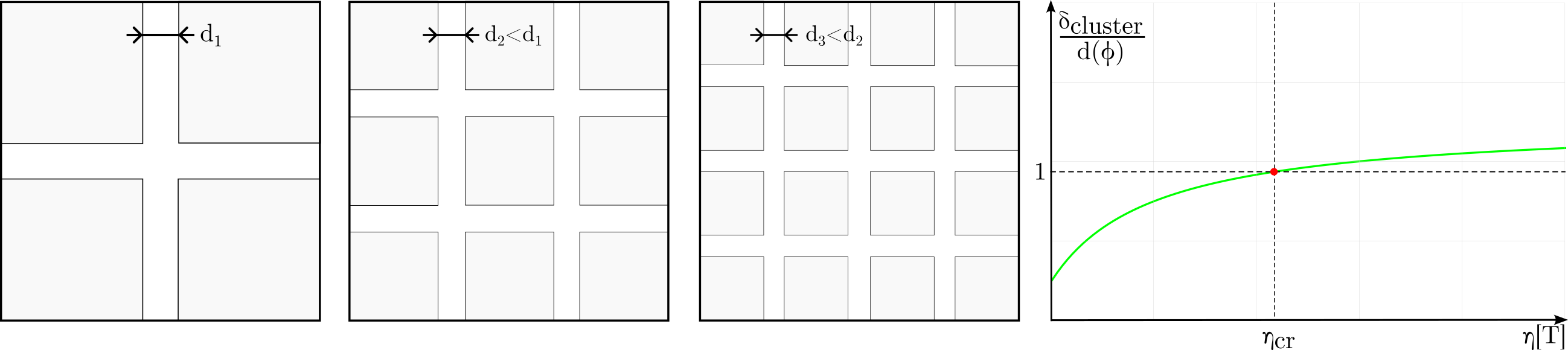}
    \caption{Three different pore channel structures (on the left) to which the same total porosity corresponds. On the right, a qualitative curve representing the pore channel occlusion that occurs at a critical value of magnetization. Increasing magnetization increases the size of the cluster $\delta_{\textup{cluster}}$ formed by the magnetic particles in the pore, until at a certain threshold $\delta_{\textup{cluster}}=d$, flow gets impeded.}
    \label{fig:enter-label}
\end{figure}
Clearly, the level of magnetization that can trigger this behaviour depends on the dimensions of the particles as well as on the porous channel characteristic size. This value, $\eta_{cr}$ henceforth, is critical in the sense that alter significantly the behaviour of the overall solid as the fluid itself cannot escape and a solid-like behaviour is shown. In this situation the poroelastic material is composed of two elastic materials where the MRF takes part in the overall elastic response by affecting the constitutive properties. 

\subsection{On the fluid viscosity} \label{ch:fluid_constitutive}
\begin{figure}[H]
    \centering
    \includegraphics[width=0.5\linewidth]{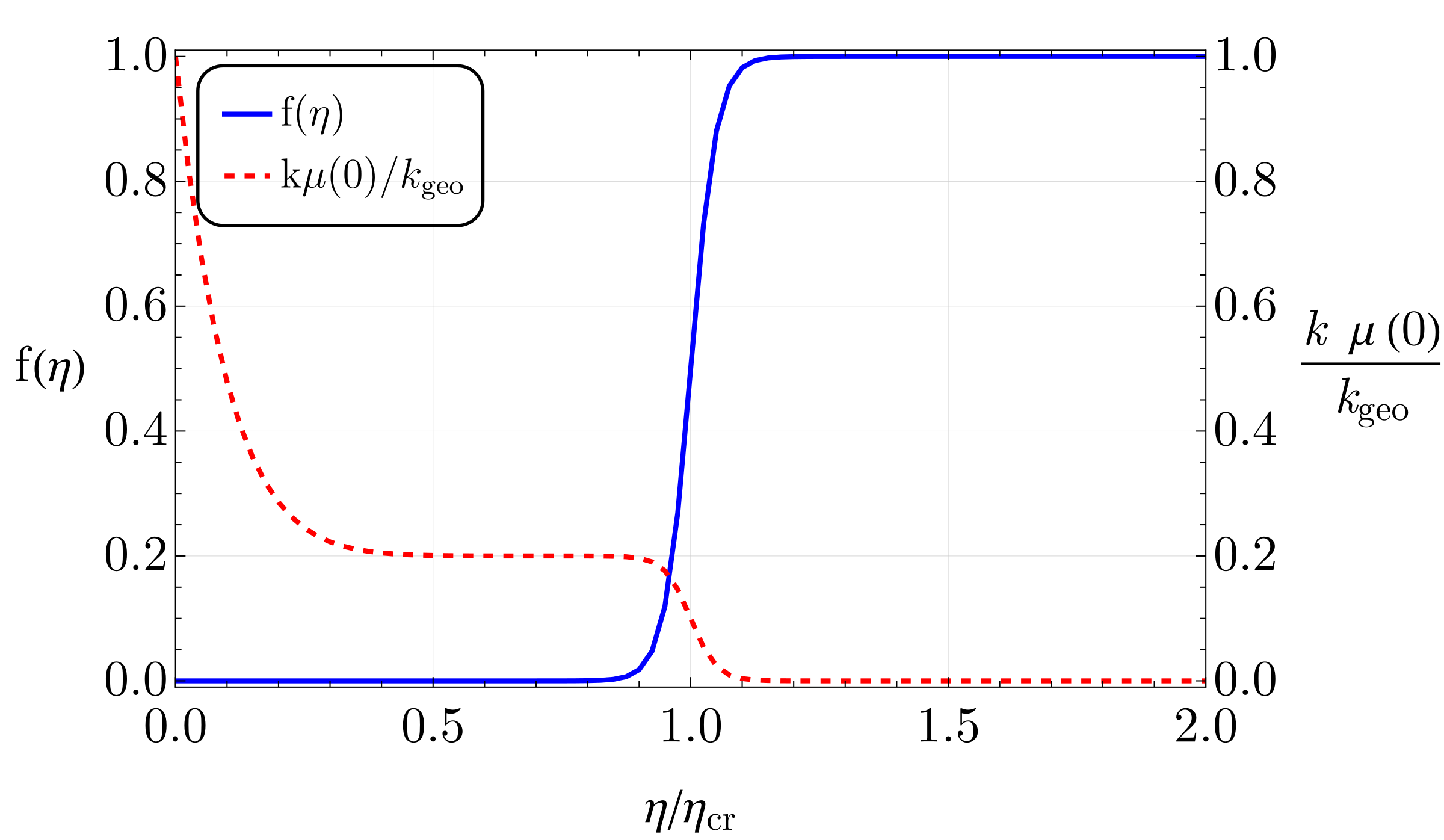}
    \caption{Graphs of the two functions $f$ and $\mu$ for varying viscosity parameter $\eta$. The plots have been obtained by choosing a representative value for $s=20$, $c=0.2/\mu(0)$.}
    \label{fig:qualitative_f_eta}
\end{figure}

The absence of a magnetic field acting on the fluid represents a state of fluid-like flow with a total viscosity exhibited by the fluid that is only due to the particle suspension in mineral oil ($\eta=0$ and $\mu=\mu(0)$).
At this stage, its effects in the poroelastic formulation are seen only through pore pressure $p$. \\
Whenever the magnetic field is turned on and the viscosity parameter ($\eta$) grows the fluid viscosity is increased accordingly resulting in a reduction of the permeability. Upon reaching the critical threshold value of the magnetic field, the structures of nano- and micro-particles inside the fluid 
become pervasive and the fluid sharply changes its mechanical response to one more similar to an elastic solid. These observations are completely captured by a Darcian-like flow in which both the viscosity $\mu$ and the permeability $k$ are functions of the parameter $\eta$ as in
\begin{equation}
    k=\frac{k_{\textup{geo}}}{\mu(\eta)}(1-f(\eta))=k_{\textup{geo}}\left[\left(\frac{1}{\mu(0)}-c\right)\left(1-\frac{\eta}{\eta_{\textup{cr}}}\right)^{s/2}+c\right]\left(1-f(\eta)\right)\quad ,
    \label{eq:permeability}
\end{equation}
where $\mu(0)$, the viscosity of the magnetorheological fluid in the absence of acting magnetic fields, $s$ and $c$ are all parameters to be fitted using the real viscosity versus magnetization experimental measurements of Fig. \ref{fig:mrf_viscous}. These functions are visible in Fig. \ref{fig:qualitative_f_eta}. To account for the introduction of the threshold in the equation of permeability (\ref{eq:permeability}), the term $(1-f)$ was introduced.

Hence, for an isotropic porosity, $\mathbf{K}=k_{\textup{geo}}\mathbf{I}$, one has:
\begin{equation*}
    \mathbf{q}=-k\ \nabla p=-\frac{k_{\textup{geo}}}{\mu(\eta)}(1-f(\eta))\nabla p\quad ;
\end{equation*}
accordingly, \cref{eq:f,eq:permeability} naturally define a complementary condition that correlates changes in mechanical properties with the fluid flux at each instant of a deformation process:
\begin{equation}
    \Delta G\cdot k=\left[G(\eta)-G(0)\right]k_{\textup{geo}}(1-f)\mu^{-1}=0\Rightarrow f(1-f)=0\quad, 
    \label{eq:complemetary}
\end{equation}
where $G$ is that of eq. \eqref{eq:constitutive_scaling}. Equation (\ref{eq:complemetary})  respects any requirement for thermodynamical consistency as well as furnishes a unified framework for the metaphysical problem involving an active fluid and a porous elastic matrix. 

\subsection{Constitutive law of the fluid}
\noindent Reformulation of (\ref{eq:poroelastic_pores}) allows to write
\begin{equation*}
    \zeta=\mathbf{A}:\mathbb{S}:\mathbf{T}^{\textup{eff}}+M^{-1}(p-p_0)\quad ,
\end{equation*}
the constitutive relationship between pressure $p$ and pore volume fraction $\zeta=\phi-\phi_{F,0}$. To completely define this relationship, the two material properties (Biot effective stress tensor $\mathbf{A}$ and Biot modulus $M$) must be evaluated. Under the hypothesis of isotropy, the Biot effective stress tensor becomes $\mathbf{A}=\alpha \mathbf{I}$, with $\alpha=1-K/K^{\textup{mat}}$, $K$ and $K^{\textup{mat}}$ being the drained and matrix bulk moduli, respectively. Hence, only two scalar constants need be defined: $\alpha$ the part of volumetric strain caused by the change of porosity when the pressure is held constant (drained condition); $M^{-1}$ the change in porosity due to a change in pressure when the strain volumetric deformation is fixed.
These parameters are typically referred to the undrained and the drained mechanical properties, so to have: 
\begin{equation*}
    \alpha=\frac{3}{B}\frac{(\nu^{(u)}-\nu)}{(1-2\nu)(1+\nu^{(u)})}  \quad , \quad M^{-1}=\frac{9(\nu^{(u)}-\nu)}{B^2E(1+\nu^{(u)})}\left(1-\frac{\nu^{(u)}-\nu}{(1-2\nu)(1+\nu^{(u)})} \right) \quad ,
\end{equation*}
where $\nu^{(u)}$ is the Poisson ratio exhibited by the overall continuum in the undrained conditions, $E$ the Young's modulus in drained conditions, and B is the trace of $\mathbf{B}=\frac{B}{3}\mathbf{I}$, called the Skempton compliance difference tensor in the isotropic case.
Hence the constitutive equation of the pore volume fraction with respect to the pore pressure becomes $\zeta=\alpha\ \textup{tr}\mathbf{E}+M^{-1}(p-p_0)$.

\paragraph{Accounting for fluid phase transition in the stress} To correctly account for the transition of the fluid to a solid-like substance it is necessary to introduce a correction factor in the definition of the stress. In fact, whenever the fluid has transitioned completely to a solid state, the pressure term $\alpha(p-p_0)$ in Eq. (\ref{eq:poroelastic_stress}) must vanish as the contribution to the stress of the interstitial substance is incorporated in the definition of the constitutive elastic tensor of Eq. (\ref{eq:overall_elastic_tensor}). To implement this logic one should write
\begin{equation}
    \mathbf{T}=(1-\phi)^{\beta}\mathbb{C}_{\textup{mat}}:\mathbf{E}+\phi^{\beta}f(\eta)\mathbb{C}_{\textup{F}}^{O}:\mathbf{E}-\alpha(1-f(\eta))(p-p_0)\mathbf{I}.
\end{equation}

\subsection{Balance of linear momentum} \label{ch:balance_solid}
At each instant of time, during a deformation process (which is here considered static for simplicity), the solid skeleton need be in equilibrium, a condition for which the following balance of linear momentum is a necessary and sufficient condition
\begin{equation}
    \nabla \cdot \mathbf{T}=\mathbf{0}\quad ,
    \label{eq:linear_momentum}
\end{equation}
which in virtue of (\ref{eq:poroelastic_stress}) and (\ref{eq:overall_elastic_tensor}) becomes:
\begin{equation*}
    \nabla \cdot ((1-\phi)^{\beta}\mathbb{C}_{\textup{mat}}:\mathbf{E}+\phi^{\beta}f(\eta) \mathbb{C}_{\textup{F}}^O:\mathbf{E}-\alpha(1-f(\eta))(p-p_0)\mathbf{I})=\mathbf{0}\quad .
\end{equation*}
\subsection{Balance of mass and fluid content} \label{ch:balance_fluid}
To solve equation (\ref{eq:linear_momentum}), the knowledge of pore pressure is necessary, as is visible from its explicit form. The pressure is regulated by the variation of pore volume fraction through the constitutive equation (\ref{eq:constitutive_pores}). Balance of mass for the pore fluid requires that
\begin{equation}
    \frac{\partial \zeta}{\partial t}+\nabla \cdot \mathbf{q}=0\quad ,
    \label{eq:fluid_content}
\end{equation}
particularized in our case as:
\begin{equation*}
    \frac{\partial }{\partial t}\left(\alpha\ \textup{tr}\mathbf{E}+M^{-1}(p-p_0)\right)-\nabla \cdot k\nabla p=0 \quad .
\end{equation*}

\section{Paradigmatic cases at comparison}
\label{result_discussion}
In this section, we present numerical solutions to a set of well-established benchmarks to evaluate the performance and accuracy of the proposed model. Each benchmark has been selected based on its relevance and ability to highlight specific characteristics of both the theory and the material. The subsections that follow provide detailed descriptions of the benchmark problems, along with their respective results and analyses. 

\subsection{The Oedometric test}
\label{ch:benchmark_1}
\begin{figure}[!ht]
    \centering
    \includegraphics[width=\linewidth]{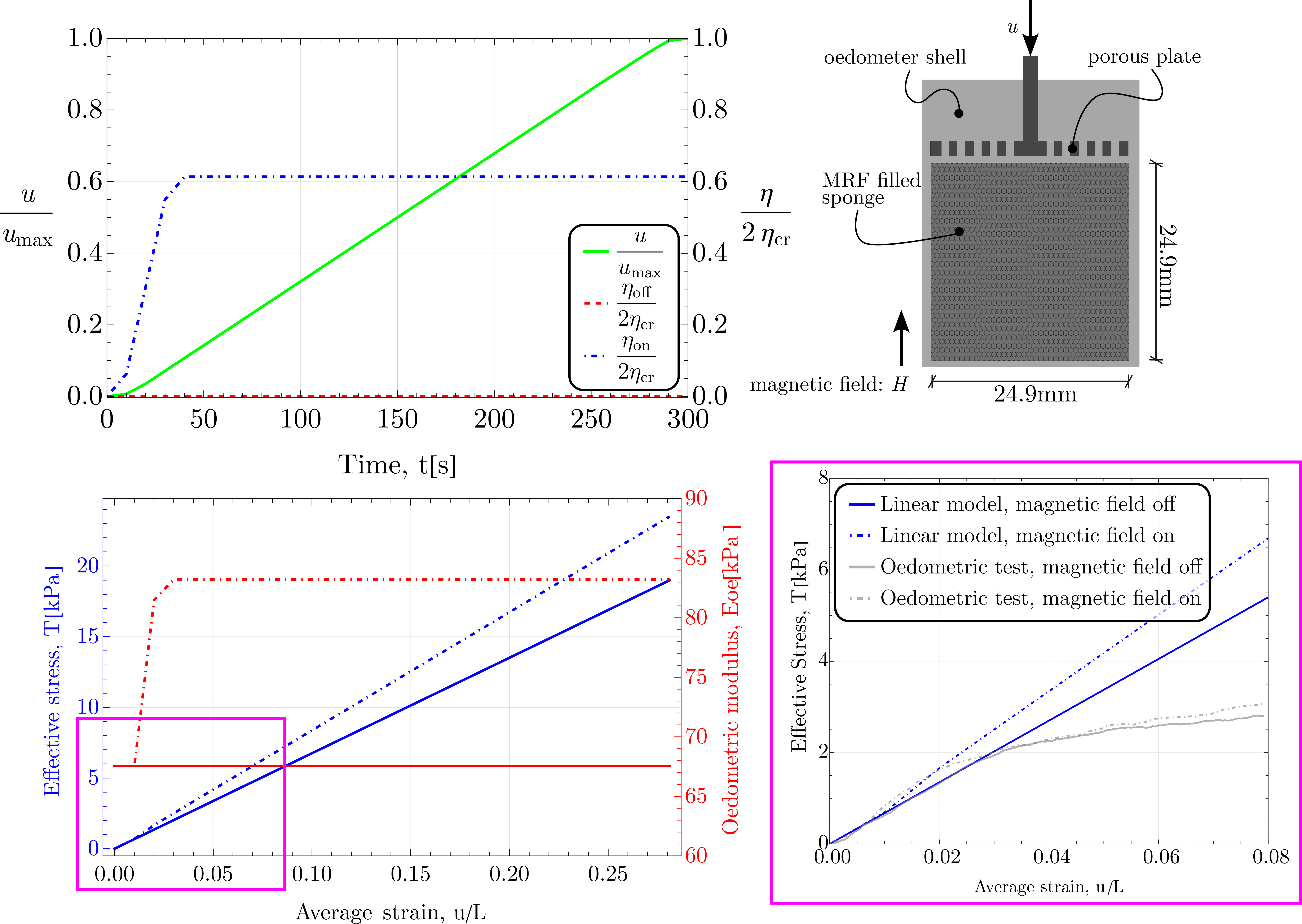}
    \caption{Results of the numerical simulation of an oedometric test carried out on a porous matrix with a magnetorheological fluid (obeying eq. \eqref{eq:complemetary}) embedded in it. A sketch of the oedometric test is depicted in the upper right part. In the upper left part: time evolution of some \textit{input} parameters, the displacement of the oedometer head ($u$) and the magnetization of fluid ($\eta$). Plots of the stress and the oedometric modulus registered during the test as the head of the oedometer is displaced downwards (lower left) with an enlargment of the graph showing comparison with experimental tests carried out in the above section. In the lower graphs, the continuous lines represent the result of the test in complete absence of magnetization -- so for an inactive fluid, while dot-dashed lines depicts the results for the magnetized case. L=24.9mm is the height of the specimen placed in the oedometer.}
    \label{fig:oedometric_test}
\end{figure}

We here specialize the group of equations regulating the poroelastic problem for the case of an oedometric test. An oedometer allows to perform under drained conditions an axial compression, by displacing the head plate, while impeding lateral displacements, see Fig. \ref{fig:oedometric_test}. The oedometric test is carried out by imposing vertical displacements (along the z-axis) while the lateral (x-axis and y-axis) are set to zero.
From the modelling point of view this results in obtaining special uniaxial loading conditions.
In particular, Eq. (\ref{eq:linear_momentum}) becomes a single scalar equation:
\begin{equation}
   \left(E_{\textup{oe}}\varepsilon-\alpha(1-f(\eta))(p-p_0)\right)_{,z}=0 \quad ,
\end{equation}
where $E_{\textup{oe}}=K+4G/3$ is the oedometric modulus, and $\varepsilon$ is the linearized deformation measure in the vertical direction. Enforcing similar test conditions on the balance of fluid mass furnishes the specialization of the second poroelastic equation:
\begin{equation}
    \left(\alpha \varepsilon+M^{-1}(p-p_0)\right)_{,t}=\left(kp_{,z}\right)_{,z}\quad .
\end{equation}
The functions ($\alpha, M, \varepsilon, p, k$) depend on a single spatial variable and on time $(z,t)$, accordingly, the partial derivatives with respect to these variables have been indicated through the subscripts \lq \lq $,z$" and \lq \lq $,t$".\\
Boundary conditions and initial conditions are expressed as: $\{u(0,t)=0,\ u(h,t)=g(t), u(z,0)=0\}$, for the displacement variable, and $\{p_{,z}(0,t)=0,\ p(z,0)=p_0\}$ for the pressure variable. Here, the function $g(t)$ represents the history of movement of the displaced head. The remaining boundary condition, necessary to solve the pressure equation, depends on whether the test is carried out in drained or undrained settings to which a Dirichlet condition or a Neumann condition on $p$ at $z=h$, corresponds respectively. In a classical way then, in the drained case the pressure must match the atmospheric one $p(h,t)=p_{\textup{out}}$, while in the undrained case, the flux must be null $q(h,t)=0$. 

Figure \ref{fig:oedometric_test} shows the results of an oedometric test carried out on the porous material described in the previous chapters, hence the mechanical parameters involved are those reported above. The test is conducted with both an inactive fluid (continuous lines) and an activated one (dashed lines). Following the experiments carried out previously, the geometry of the sample and the loading path are chosen accordingly. The test is thus a good example of how it is possible to change the overall mechanical and hydraulic properties of the porous solid, obtaining a stiffer or less stiff material with time-regulated response.

\subsection{Three-point bending test}
\label{ch:benchmark_2}
\begin{figure}[!h]
    \centering
    \includegraphics[width=0.8\linewidth]{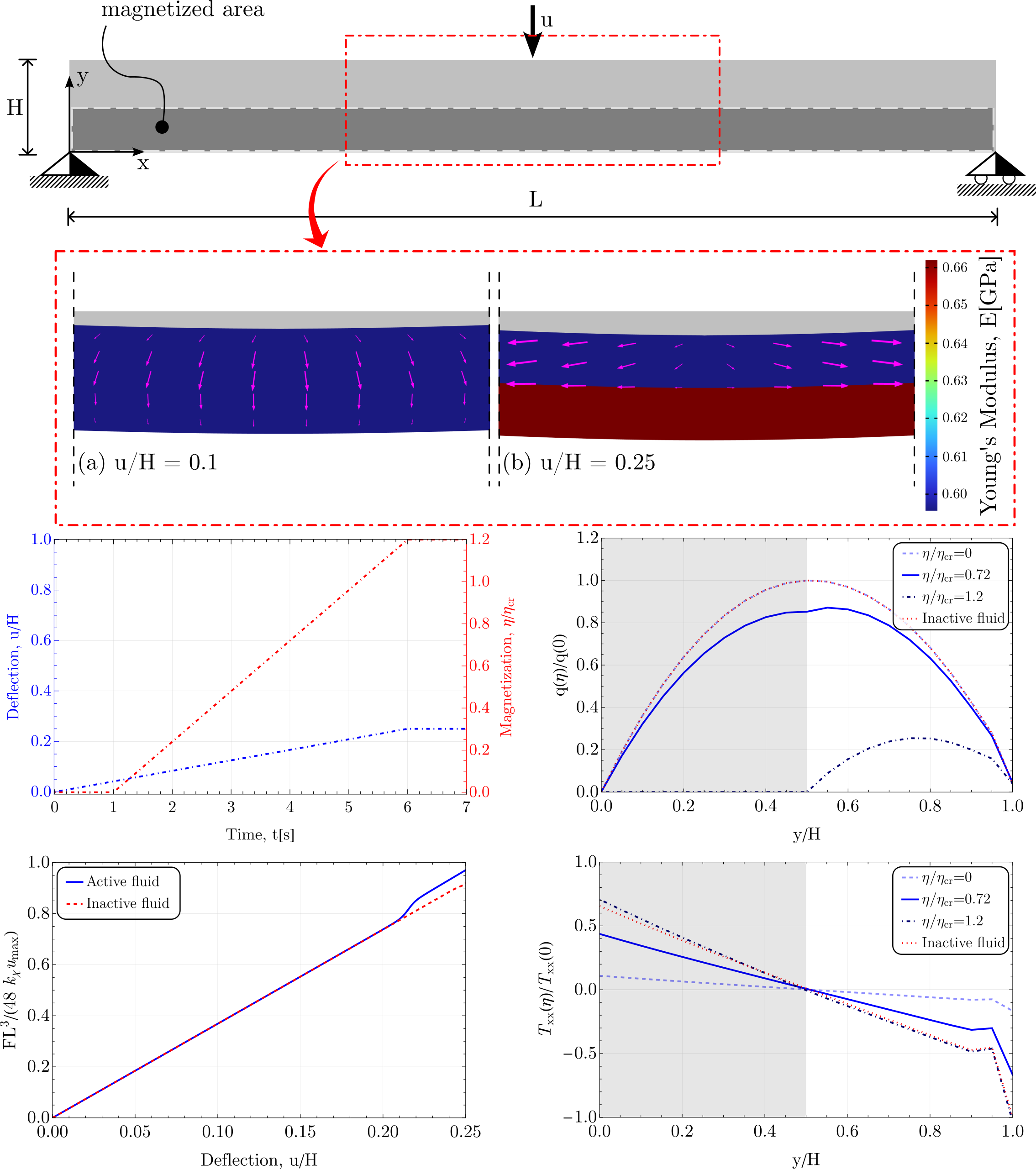}
    \caption{In the upper part, the sketch of the three-point bending test carried out on a poroelastic beam with a magnetorheological fluid embedded within. The lower half of the beam is magnetized during the test and the result in terms of fluxes and stresses are plotted in the middle and lower part of the panel. As the displacement is induced the beam at first (a) behaves as a normal poroelastic one with fluxes pointing from the compressed regions to the tensioned ones. Whenever magnetization is induced (b) fluxes are impeded in the lower region of the beam as shown by the purple arrows. Likewise, Young's Modulus is changed due to the participation of the phase-transitioned fluid in the constitutive elastic formulation -- according to (\ref{eq:overall_elastic_tensor}). In the middle part of the panel, on the left, the input displacement and magnetization are reported, showing a piecewise monotonic increase for both. On the right side, the vertical component of the flows in the middle cross-section of the beam for three different levels of magnetization and for the case of poroelastic beam with inactive fluid (dotted in red). Lastly, the lower graphs show the force versus displacement plot (on the left) and the horizontal normal stresses in the middle-span cross-section for various levels of magnetization.}
    \label{fig:3PB}
\end{figure}

With rehabilitation-oriented applications in mind, and in particular with reference to possible functionalized patches and bandages that require delivering drugs to specific tissues over prolonged periods, we here explore the model's ability to treat bending of thin beams and spatial tunability of mechanical parameters. As a concluding benchmark, therefore, we present the result of a numerical three-point bending test carried out on a poroelastic beam under plane stress conditions. As explained above the porous beam is pervaded by a magnetorheological fluid whose properties are controlled by activating a magnetic field during the test. The beam of height H and length L, is tested by imposing a vertical displacement at half its length while forces are recorded. Amidst this monotonic loading process (at t=1 second), the lower half of the beam is hit with a magnetic field which slowly changes the rheological properties of the fluid, see Figure \ref{fig:3PB} plot of the deflection and magnetization as a function of time (middle-left part of the panel). Magnetization is progressively increased until phase-transition occurs at a critical value of $\eta$ when the fluid can not escape anymore due to clusters of magnetic particles obstructing the flow. When this happens, several things change from the static point of view for the beam: (i) the lower half stiffens up following (\ref{eq:overall_elastic_tensor}) (also visible from the colour variation of the (overall) Young's modulus, E, in upper part of Figure \ref{fig:3PB}) causing an increase in the force registered (see Figure \ref{fig:3PB}, lower-left plot); (ii) fluxes (purple arrows in upper part of Figure \ref{fig:3PB}) are completely impeded in the lower half portion of the beam and the pressure and flux profile change drastically (see Figure \ref{fig:3PB}, middle-right plot for the profile of the vertical flux in the middle cross-section of the beam). 

The force versus displacement plot has been normalized with respect to the force needed to deflect a linear elastic beam with the overall elastic properties proper of the poroelastic beam under study with full magnetization for which the bending stiffness is easily evaluated as: $$k_{\chi}=I\ \left[E(f=0)^2+14E(f=0)E(f=1)+E(f=1)^2\right]/8(E(f=0)E(f=1)))\ ,$$ $I$ being the moment of inertia of the cross-section with respect to an axis positioned at half the section height, while $E(f=0)$ and $E(f=1)$ are the overall Young's modulus of the material when magnetization is off and on, respectively. In the absence of comparison real experiments, the current numerical simulation has been carried out with an isotropic matrix defined by $E=0.6$GPa, $\nu=0.25$. \\

\begin{figure}
    \centering
    \includegraphics[width=0.8\linewidth]{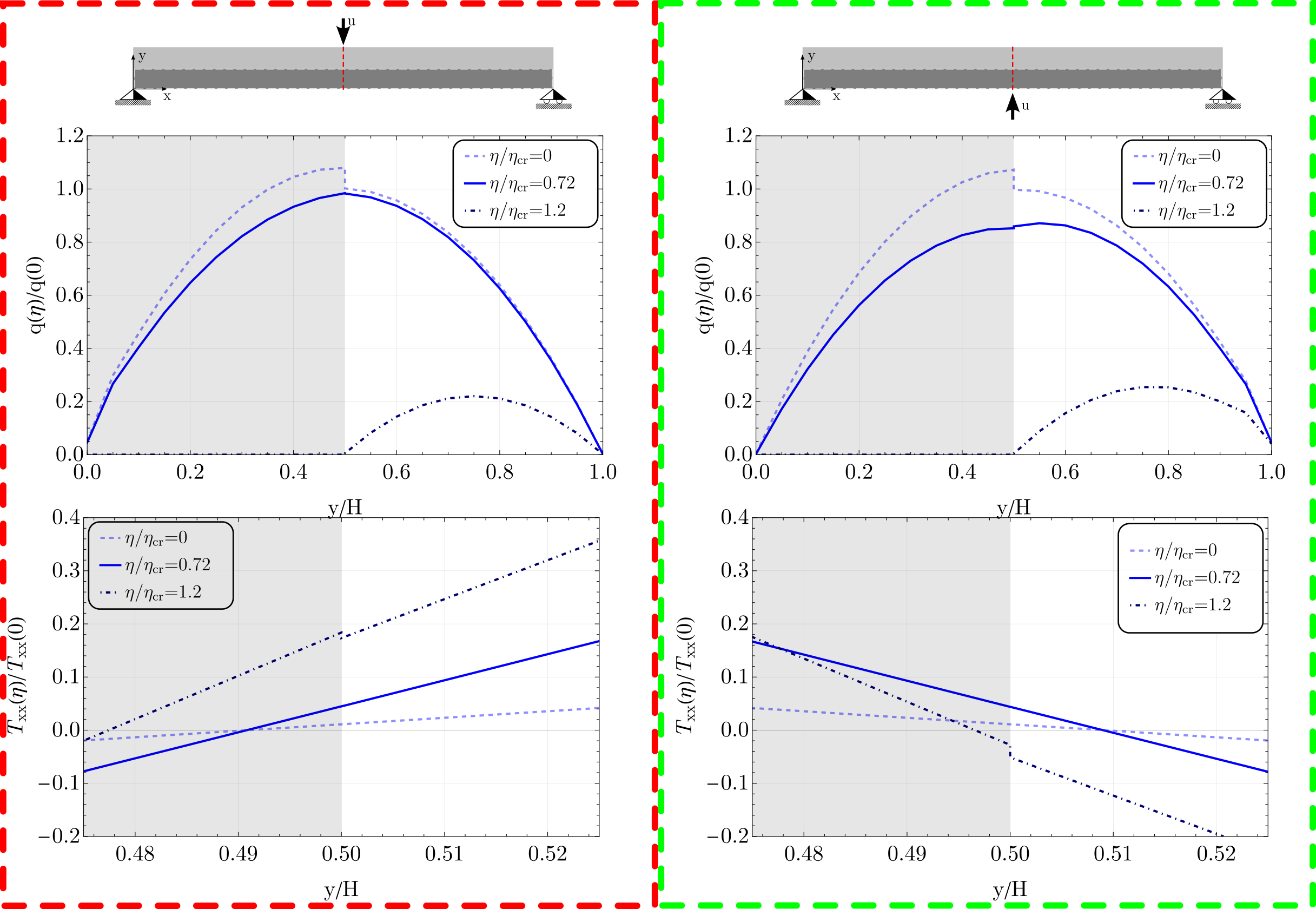}
    \caption{Sensitivity of a beam made of the proposed composite porous material to the direction of the applied displacement in a three-point bending test. On the left, the beam is deflected downward, while on the right it is deflected upward. This inversion has no effect on the mechanical response when the magnetic field is turned off. However, once the magnetic field is activated and the lower half of the beam becomes fully magnetized, the vertical flow patterns across the mid-width section (first row of plots) clearly differ between the two loading directions. Moreover, the axial stress component, $T_{xx}$, is not simply the mirror image of the downward case with respect to the mid-plane along the y-direction.}
    \label{fig:3PB_inverted}
\end{figure}

\noindent Interestingly, the response of the beam to deflection appears to be non-symmetric with respect to the load direction, see Figure \ref{fig:3PB_inverted}. Indeed, changing the direction of loading shows non-trivial consequences on the cross-sectional distribution of horizontal normal stress, $T_{xx}$, and on fluxes, a fact which is ascribable to the repositioning of the neutral axe (the axe parallel to the main dimension of the beam that is the loci of all points of the continuum with null horizontal normal stress $T_{xx}$ in a pure bending context \cite{krank2016}) as magnetization occurs. In the gravity-oriented set-up of the load (see Figure \ref{fig:3PB_inverted}) on the left, in fact, the axe lies in the lower half of the beam (where magnetization occurs) throughout the whole test so that fluxes have a monotonic trend in both lower and upper part of the beam. On the contrary, in the counter-gravity-oriented (see Figure \ref{fig:3PB_inverted} on the right), the neutral axe shifts its position during the test by crossing the interface between the magnetized and non-magnetized region inducing a non-monotonic profile of the flow in both of them. This condition slightly highlights the asymmetry of the set-up to loading conditions and opens up various possibilities for the employment of such structure in several engineering fields, such as that of rehabilitation of human tissues where control of the fluxes can lead to engineered patches. 

\section{Conclusions}
The demand for more effective rehabilitation solutions has led to the exploration of lightweight, adaptable exosuits. The combination of porous media and MR fluids offers a promising avenue for achieving the desired balance of comfort, adaptability, and support. Propelled by the need of designing next-generations components for engineering at large implementing the logic of combined MR fluids and porous deformable medium a modeling approach which lay its foundation on sound and solid mechanical concepts are required for the correct design and optimization procedures.\\ 
In this framework the present work explores the possibility of modeling such complex active systems using the poroelasticity theory of Biot and colleagues. As a first attempt the linear regime is developed accounting for a porous matrix with certain fixed porosity and an active fluid whose viscosity evolves following magnetization. Phase transition of the MRF is uploaded in the model in a mixture theory fashion. The model is then tested under well-known benchmark conditions typical of poroelastic materials, namely the oedometric test and a three-point bending test to highlight several properties of the technology proposed that meet several requirements for rehabilitation-related applications. The oedoemtric test is compared also with experimental tests carried out using ad hoc developed apparatus, showing great correspondence.\\
In this context, the proposed model represents a significant step forward in the endeavour of providing a foundation for the development of next-generation exosuits that can be tailored to meet the specific needs of each patient, ultimately enhancing the rehabilitation process and improving outcomes.

\section*{Acknowledgements}
All the Authors thank the financial support from the project FIT4MEDROB, PNC0000007 (ID 62053).
M.F. additionally thanks financial support from MUR through the project AMPHYBIA (PRIN-2022ATZCJN).
A.C. has been supported by the Project of National Relevance PRIN2022 grant no. P2022XLBLRX and PRIN2022PNRR grant no. P2022MXCJ2, funded by the Italian Ministry of University and Research (MUR).

\bibliographystyle{elsarticle-num}
\bibliography{reference}

\end{document}